# Extended opportunity cost model to find near equilibrium electricity prices under non-convexities

Hassan Shavandi, Mehrdad Pirnia, J. David Fuller

*Department of Management Sciences, University of Waterloo, Waterloo, Canada*

## Abstract

This paper finds near equilibrium prices for electricity markets with non-convexities due to binary variables, in order to reduce the market participants' opportunity costs, such as generators' unrecovered costs. The opportunity cost is defined as the difference between the profit when the instructions of the market operator are followed and when the market participants can freely make their own decisions based on the market prices. We use the minimum complementarity approximation to the minimum total opportunity cost (MTOC) model, from previous research, with tests on a much more realistic unit commitment (UC) model than in previous research, including features such as reserve requirements, ramping constraints, and minimum-up and -down times. The developed model incorporates flexible price-responsive demand, as in previous research, but since not all demand is price responsive, we consider the more realistic case that total demand is a mixture of fixed and flexible. Another improvement over previous MTOC research is computational: whereas the previous research had nonconvex terms among the objective function's continuous variables, we convert the objective to an equivalent form that contains only linear and convex quadratic terms in the continuous variables, thus allowing for efficient optimization by CPLEX.
We compare the unit commitment model with the standard social welfare optimization version of UC, in a series of sensitivity analyses, varying flexible demand to represent varying degrees of future penetration of electric vehicles and smart appliances, different ratios of generation availability, and different values of transmission line capacities to consider possible congestions.  The minimum total opportunity cost and social welfare solutions are mostly very close in different scenarios, except in some extreme cases; the obtained solution has smaller opportunity costs than social welfare, at very little cost in reduction of social welfare; and solution times of the minimum total opportunity cost model are longer, due to a larger model size, but the times are still very quick for practical use.

## Keywords

# 1. Introduction

In many electricity systems that are run as markets, the Independent System Operator (ISO) uses the solution of a mixed integer problem (MIP), such as the Unit Commitment (UC) problem, to determine the "dispatch" instructions to generators, as well as the electricity market prices Steeger and Rebennack (2015). Typically, the MIP minimizes the total cost of generation over the next 24 hours, in order to meet the forecast demand, which is assumed to be fixed and unresponsive to prices. If demand is responsive to price, then the minimum cost objective is replaced by an objective which maximizes social welfare (SW), defined as the value to consumers of the purchased electricity (measured as area under the demand curve), minus total costs of generation; sometimes the minimum cost, fixed demand model is also called the SW model. However, due in part to the presence of non-convexities in the UC model (usually in the form of binary variables that are used to represent, e.g., startup/shutdown decisions, minimum supply requirements, and indivisibilities) the linear prices obtained from such mixed integer problems cannot guarantee satisfactory profit to generators to recover their cost of generation (Scarf, 1994). The most glaringly unsatisfactory profit is an accounting loss (revenue is less than costs incurred by following the ISO's instructions), but sometimes even a positive profit is unsatisfactory to the generator if, at the prices announced by the ISO, the generator could theoretically earn a larger profit by generating and selling a quantity of electricity that is different from the ISO's instructions. Whether the unsatisfactory profit is negative or positive, we can say that the generator incurs an opportunity cost. To avoid such situations, other pricing schemes are proposed in the literature and some are used in practice to complement or modify the prices obtained directly from mixed integer programs.

One of the main corrective instruments, practiced in some jurisdictions (e.g., PJM) is the make-whole payment, where market participants are paid a fixed lump sum to make up for their negative profit (Bresler, 2014). These make-whole payments are funded by special lump sum, fixed charges on consumers' bills, or through additions to the energy price that consumers see; in this paper, we consider a method which includes both the lump sum form of consumer charges (which fund the fixed payments to producers) and adjustments to prices which are seen by consumers and by producers. Consumers are also charged for other services provided by generators (e.g., reactive power support) and the total of all lump sum payments is called "uplift." Although uplift payments provide financial satisfaction to generators, they distort the price signals needed for market efficiency, and therefore market operators try to reduce total uplift payments and to favor price-based consumer charges and compensation of producers.

In O'Neill et al. (2005) the linear prices in a MIP model of a market are obtained from the dual variables of market clearing constraints of an LP which is reformulated from the MIP by fixing the integer variables at their optimal values, and fixed payments are obtained from the dual variables of the constraints fixing the integer variables. Bjørndal and Jörnsten (2008) enhance the approach suggested in O'Neill et al. (2005) by adding extra constraints to the LP problem, that fix some of the binary and continuous variables. Hogan and Ring (2003) and Gribik, Hogan, and Pope (2007) retain the optimal primal solution of the MIP but they obtain prices which minimize the total "uplift" which is defined as the total of all generators' opportunity costs, relative to their profits from following the generation quantities from the MIP. Hua and Baldick (2016) devise a computationally efficient way to calculate these minimum uplift prices, called "convex hull pricing"; they note that the Midcontinent ISO in the USA has used an approximation of convex hull pricing. Galiana, Motto and Bouffard (2003) also retain the optimal primal solution to the MIP, and they suggest an approach to reduce the uplift payments to zero by adjusting prices and by taking lump sum payments from some producers and giving them to others, to remove opportunity costs. In Araoz and Jornsten (2011), the price is determined by formulating a semi-lagrangian relaxation of the MIP problem, and using its dual prices.

In a recent study, Liberopoulos and Andrianesis (2016) provide a critical review on current pricing schemes and propose a new pricing approach called minimum zero-sum uplift. Zoltowska (2016a) discusses a method to eliminate uplift, while keeping the MIP primal solution, by adjusting consumer prices (but not producer prices) upward, to cover the lump sum payments to producers. Zoltowska (2016b) extends this method to permit the shifting of demand to different hours of the day.

The methods suggested in the above literature assume a non-price-responsive demand function, and all of them retain the primal MIP solution from the SW model. However, considering the advances in smart grid technologies, more and more demand will be able to respond to market price signals; recent research that incorporates price responsive demand in UC models includes, e.g., Tumuluru et al. (2017), Jonghe et al. (2017), Wu (2013) and Zoltowska (2016b). Therefore, fixing the supply and demand quantities obtained from the SW model and then adjusting prices is not an appropriate approach to find market prices.

In Gabriel et al. (2013b,c), a method is suggested to find prices, but with quantities that are not obtained from the SW problem. In Gabriel et al. (2013b,c), first a mixed complementarity problem (MCP) is constructed (Gabriel et al. 2013a) by relaxing the binary variables, and deriving Karush-Kuhn-Tacker (KKT) conditions, while considering market clearing constraints and price-responsive demand functions. The discrete conditions are finally imposed into the MCP, and the discretely constrained MCP is solved by minimizing the sum of the complementarities, while the non-complementarity constraints are respected. Although the proposed method provides near equilibrium market prices, and can be applied when price responsive demand exists, the notion of the minimum complementarity (MC) does not have a practical meaning for electricity markets. Gabriel (2017) proposes a different method to solve the discretely constrained MCP, but the relation to economic thinking is not explored.

Abbaspourtorbati et al. (2017) propose a pricing procedure that is somewhat similar to that of Gabriel et al. (2013b,c), but for a fixed demand setting: they relax the binary variables of the UC MIP, giving a linear program (LP), they formulate the dual LP, and finally they create a new MIP with all primal and dual constraints (including binary primal variables), and an objective which minimizes the primal objective minus the dual objective.

Recently, Huppmann and Siddiqui (2018) proposed a general method to solve for Nash equilibrium in games with continuous and binary variables, by including compensation payments for opportunity costs directly in an optimization model. They illustrate the method with a simplified power market uplift problem with variable demands (bounded above) and fixed marginal values of the demands (i.e., horizontal demand curves).

Fuller and Çelebi (2017) extend the minimum uplift approach that began with Hogan and Ring (2003), to allow for price responsive demand. The initially defined Minimum Total Opportunity Cost (MTOC) minimizes the total opportunity costs of all generators, but the solution to the MTOC model is extremely difficult to compute. However, Fuller and Çelebi (2017) show that the minimum complementarity approach of Gabriel et al. (2013b,c) to solving the discretely constrained MCP provides an approximation of the solution of the MTOC model, thus providing a practical meaning of the minimization of complementarity, as well as a computational method to approximate the solution of the MTOC model. We refer to the MTOC model, solved by the minimum complementarity approach as the MTOC-MC model. Fuller and Çelebi (2017) illustrate the MTOC-MC model with a tiny, one-period UC model without a transmission network, and with a two period UC model with a small, six-bus transmission network.

In this paper, we want to investigate the practicality of the theories suggested in MTOC-MC model, by extending it to an applied UC model with realistic constraints on a larger test system. In the course of developing and solving the model, we made the following specific contributions:

- in addition to the features in the largest UC model tested in Fuller & Çelebi (2017), we have also included realistic physical constraints of unit commitment models, such as: reserve requirements expressed as a fraction of demand, ramping constraints and costs, minimum-up and -down times, and demand as a *mixture* of price-responsive and fixed demand (we also vary the mixture to simulate different degrees of penetration of automatic equipment that allows consumers to respond to prices)
- the MINLP form of MTOC-MC, with nonconvex objective terms such as price times generation amount, is converted to MIQCP form, with only linear and convex terms in the objective, allowing for the use of CPLEX, for efficient optimization calculations

- we conduct tests on realistic data, with a much larger system than in Fuller and Çelebi (2017) -- 24 buses versus 6 buses, and 24 hours versus just two hours in Fuller and Çelebi (2017)
- we conduct comprehensive sensitivity analysis to investigate the practicality of this model in real applications comparing to more traditional models, and evaluate the financial performance of players in the market by varying flexible demand, generation and transmission line capacities.

Some of the interesting observations from the tests are as follows:

- MTOC-MC is of course larger than SW, and therefore takes longer to solve, but is still quick enough to be useful
- MTOC-MC reduces opportunity costs compared to SW, as expected, but it is somewhat surprising that this reduction comes at very little cost in reduction of social welfare
- SW and MTOC-MC solutions are close but compared to SW, MTOC-MC has larger generators' profit and smaller total opportunity cost, and these differences are especially large when demand is mostly fixed (i.e., very little price-responsiveness), or the network is highly congested.

The rest of the paper is organized as follows: section two begins with formulating the model of consumers and follows by formulating the optimization problems of consumers, market participants and the SW model. The third section presents the formulation of the MTOC-MC model. In section four, the proposed model is tested and validated on the IEEE RTS 24-Bus test system. Finally, in section five, directions for future research are suggested.

## 2. The optimization problems of consumers, market participants and the SW model

This section presents the SW UC model. In section 2.1, the model of consumers is formulated using a linear marginal value function of demand, and a minimum demand quantity. Secondly, market entities, including generation units and transmission network operator are individually modeled in sections 2.2 and 2.3, in which they maximize their profit. Finally, in section 2.4, all the individual models are used in order to derive the SW model. The individual models are also used to build the MTOC model in section 3. Appendix A defines all the notations used to formulate the above problems.

### 2.1. Model of Consumers

In most day-ahead UC models, hourly demand forecasts are considered as fixed and unresponsive to the real-time prices, because consumers do not have the instantaneous information on prices or the capability to respond to the prices by changing their demands.

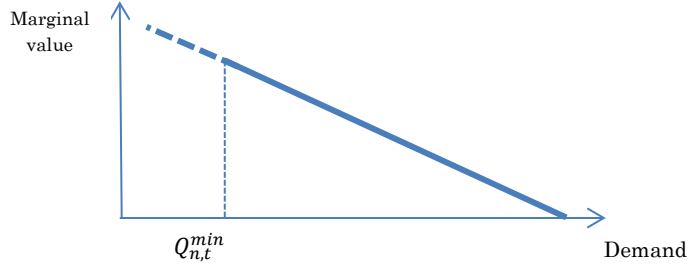

Figure 1. The linear marginal value function and minimum demand

However, economic theory suggests that there would be great benefits to flexible, price-responsive demand. Time of use pricing (see, e.g., Wang et al., 2015) is in widespread use, as a step in this direction, but full real-time pricing, the ideal, may become a reality if technologies are developed to deliver real-time price information to consumers and to respond automatically, e.g., by delaying the use of high demand devices when the price is high. We represent the partial introduction of such smart grid communication and control devices for consumers in a simple aggregate way. First, we assume that consumers at bus $n$ in hour $t$ have a linear marginal value function for electric energy, as shown in Figure 1. Second, we additionally assume that consumers are unable to respond to different real-time prices for some minimum amount of demand, $Q_{n,t}^{min}$, while for consumption above this minimum, consumers do respond to real-time prices by choosing to purchase an amount $q_{n,t}$ corresponding to the price on the marginal value curve. In our numerical illustrations, we vary $Q_{n,t}^{min}$ while keeping the marginal value curve constant, in order to represent varying degrees of penetration of smart grid communication and control in consumption. The linear relationship between marginal value and demand is formulated as shown in (1a):

$Marginal\ value = a_{n,t} - b_{n,t} q_{n,t}$ (1a)

Therefore, the total value, or benefit function is the area underneath the marginal value curve:

$$B(q_{n,t}) = \int_0^{q_{n,t}} (a_{n,t} - b_{n,t} q'_{n,t})\, dq'_{n,t} = a_{n,t} q_{n,t} - 0.5 b_{n,t} q_{n,t}^2$$ (1b)

We make the standard assumption that consumers choose an amount of consumption so as to maximize their net benefit, i.e., consumers' surplus. We further assume that consumers pay for both electric energy and for reserve capacities, in a combined pricing mechanism that is charged for every unit of consumed electric energy. The reason for this combined pricing mechanism is our assumption that the system-wide reserve requirement for hour $t$ is expressed as a fraction $\varphi$ of total demand $\sum_n q_{n,t}$ and those consumers at bus $n$ pay for reserves in proportion to their contribution to the total reserve requirement. Thus, the combined energy and reserve payment by consumers is $(p_{n,t}+\varphi p_t^R)q_{n,t}$, where $p_{n,t}$ is the electricity market price and $p_t^R$ is the market price for reserves. The surplus-maximizing behavior of consumers at bus $n$ is given by the following optimization problem (dual variables for minimum demand constraints are in square brackets).

$$\min \sum_t (p_{n,t}+\varphi p_t^R)q_{n,t} - (a_{n,t}q_{n,t} - 0.5 b_{n,t} q_{n,t}^2) \tag{1c}$$
s.t.

$$q_{n,t} \geq Q_{n,t}^{min}, \quad \forall t \qquad [\beta_{n,t}] \tag{1d}$$

## 2.2. The Optimization Model of Generation Units

In this section, a profit maximization model, based on Bergh et al. (2016) is described, to represent generators' behavior, in which all the operational constraints, such as ramping, generation capacity, minimum up and down time, and initial status of generators are considered. The given model considers two types of generators:

1. generators which can supply both energy and reserve capacity in any combination within their capacity limits; and
2. generators which can supply energy only, but not reserve capacity.

This classification of generators' reserve capabilities is very simple – the generator model in this section allows the system to respond to sudden needs for more power, due to unexpected increases in demand or the full or partial loss of a generation unit. A more complex classification and model would deal with sudden drops in demand or increases in supply from variable sources such as wind, with generators' reserve capabilities distinguished in part by how quickly they can respond (e.g., 10 minute or 30 minute spinning reserves, and non-spinning reserves).

We assume that each generator takes the price of electricity at each bus and time period, and the price of reserves at each period as given. The profit of generator $i$ is the revenue from selling energy and reserve capacity, minus the associated costs, which are startup, shutdown, ramping, and variable generation and reserve costs. The following expression shows the objective function for generator $i$ which is the negative of the generator's profit:

$$\min \sum_t \left[ \lambda_{i,t} g_{i,t} + \lambda_{i,t}^{RES} A_i^{RES} g_{i,t}^R + C_{i,t}^{SU} v_{i,t} + C_{i,t}^{SD} w_{i,t} + C_{i,t}^{RP} r_{i,t} - \sum_n (A_{i,n}^{gen} p_{n,t} g_{i,t}) \right.$$
$$\left. - p_t^R A_i^{RES} g_{i,t}^R \right] \qquad (2a)$$

The following inequalities (2b)-(2c) show ramping constraints, in which the rate of change of generation will be limited by the generators' ramping rate. The dual variables (in square brackets) are used later to derive the KKT conditions of the LP in which binary variables are relaxed:

$$g_{i,t} - g_{i,t-1} - r_{i,t} \leq 0, \quad \forall i, \forall t \qquad [\alpha_{i,t}^{RP1}] \qquad (2b)$$
$$g_{i,t-1} - g_{i,t} - r_{i,t} \leq 0, \quad \forall i, \forall t \qquad [\alpha_{i,t}^{RP2}] \qquad (2c)$$

The following inequality constraints (2d)-(2g) consider the ramping limits, and the minimum and maximum generation limits:

$$g_{i,t} - g_{i,t-1} - RU_i z_{i,t} - (SU_i - RU_i) v_{i,t} \leq 0, \quad \forall i, \forall t \qquad [\alpha_{i,t}^{RP3}] \qquad (2d)$$
$$g_{i,t-1} - g_{i,t} - RD_i z_{i,t-1} - (SD_i - RD_i) w_{i,t} \leq 0, \quad \forall i, \forall t \qquad [\alpha_{i,t}^{RP4}] \qquad (2e)$$
$$G_i^{min} z_{i,t} - g_{i,t} \leq 0, \quad \forall i, \forall t \qquad [\alpha_{i,t}^{min}] \qquad (2f)$$
$$g_{i,t} + A_i^{RES} g_{i,t}^R - G_i^{max} z_{i,t} \leq 0, \quad \forall i, \forall t \qquad [\alpha_{i,t}^{max}] \qquad (2g)$$

Also, minimum up and down time constraints are modeled as shown in (2h)-(2i):

$$D_i^{min} w_{i,t} - \sum_{k=t}^{t+D_i^{min}-1} (1 - z_{i,k}) \leq 0, \quad \forall i, \forall t \qquad [\alpha_{i,t}^{DT}] \qquad (2h)$$
$$U_i^{min} v_{i,t} - \sum_{k=t}^{t+U_i^{min}-1} z_{i,k} \leq 0, \quad \forall i, \quad \forall t \qquad [\alpha_{i,t}^{UT}] \qquad (2i)$$

In addition to the above constraints, the following constraint controls the logical status of generation units, in order to ensure that the generators' dispatch decision is aligned with the startup and shutdown decisions:

$$z_{i,t-1} - z_{i,t} + v_{i,t} - w_{i,t} = 0, \quad \forall i, \forall t \qquad [\alpha_{i,t}^{OP}] \qquad (2j)$$

Moreover, to enforce the minimum up and down times for startups and shutdowns in the previous day, the following constraints are added:

$$z_{i,t} = 1, \quad i \in I_{ON}, 1 \leq t \leq t_i^{ON} \qquad [\alpha_{i,t}^{ON}] \qquad (2k)$$
$$z_{i,t} = 0, \quad i \in I_{OFF}, 1 \leq t \leq t_i^{OFF} \qquad [\alpha_{i,t}^{OFF}] \qquad (2l)$$

Constraints (2m)-(2n) are also added to take into consideration the initial status of generators:

$$z_{i,0} - Z_{i,0}^{PT} = 0 \, , \forall i \tag{2m}$$
$$g_{i,0} - G_{i,0}^{PT} = 0 \, , \quad \forall i \tag{2n}$$

Finally, the following constraints enforce the binary and non-negativity conditions on the variables:

$$r_{i,t} \geq 0 \, , g_{i,t} \geq 0 \, , z_{i,t}, v_{i,t}, w_{i,t} \in \{0,1\}, \quad \forall i, \forall t \tag{2o}$$

The MTOC-MC and SW models, which are defined below, include all of the variables and constraints of the market participants' models, such as the binary variables $z_{i,t}$, $v_{i,t}$ and $w_{i,t}$. In order to solve the MTOC-MC and SW models efficiently, we reduce the number of binary variables by two thirds by replacing $v_{i,t}$ and $w_{i,t}$ with real nonnegative variables, at the slight expense of some additional constraints, as proposed in Hedman et al. (2010). Although this modification justifies the elimination of these binary variables only for the cost minimization model of the generators (2a)-(2o), the same arguments can be made to eliminate these binary variables from the MTOC-MC and SW models.

We define an *altered generation model* with objective function (2a), constraints (2b) to (2n), but replacement of (2o) and inclusion of new constraints as follows:

$$r_{i,t} \geq 0 \, , g_{i,t} \geq 0 \, , v_{i,t} \geq 0 \, , w_{i,t} \geq 0 \, , z_{i,t} \in \{0,1\}, \quad \forall i, \forall t \tag{2p}$$
$$z_{i,t-1} + v_{i,t} \leq 1, \quad \forall i, \forall t \qquad [\sigma_{i,t}] \tag{2q}$$

### 2.3. Transmission network operator optimization model

In this section, a standard optimization model for transmission network operator is given, in which the operator maximizes its profit through arbitrage by purchasing electricity at some buses, moving it through the network, and selling it at other buses where the price is higher. Although in many jurisdictions the transmission network is publicly owned and does not participate in the market, in this research, we have considered it as a market participant (see Gabriel et al., 2013a).

The objective function of the transmission network operator is as follows (electricity prices are taken as given parameters):

$$\min \sum_{t=1}^{T} \sum_{l=1}^{L} \sum_{n=1}^{N} p_{n,t} A_{l,n}^{grid} f_{l,t} \tag{3a}$$

where the matrix $A_{l,n}^{grid}$ indicates the connections between buses $n$ and lines $l$ using elements with value 1 if $n$ is the origin bus of line $l$, -1 if $n$ is the destination bus of line $l$, and 0 if line $l$ is not connected to bus $n$. Note that power can flow in either direction on a line; the terms "origin" and "destination" are set arbitrarily for each line in order to specify that $f_{l,t} > 0$ means flow from origin to destination, while $f_{l,t} < 0$ means flow in the opposite direction. Since the profit is made by subtracting origin price from the destination price for each unit of electricity, minimizing the objective function (3a) will maximize the profit of the transmission line operator.

The following constraints state the limits of transmission line flows (3b)-(3c) and the physical constraints of the DC power flow approximation (3d).

$$f_{l,t} - F_l^{max} \leq 0 , \quad \forall l, \forall t \qquad [\gamma_{l,t}^{max}] \qquad (3b)$$

$$-f_{l,t} - F_l^{max} \leq 0 , \quad \forall l, \forall t \qquad [\gamma_{l,t}^{min}] \qquad (3c)$$

$$B_l \sum_{n=1}^{N} A_{l,n}^{grid} \theta_{n,t} - f_{l,t} = 0 , \quad \forall l, \forall t \qquad [\gamma_{l,t}^{phys}] \qquad (3d)$$

$$-\bar{\theta} \leq \theta_{n,t} \leq \bar{\theta} , \quad \forall n, \forall t \qquad [\gamma_{n,t}^{\theta max}, \gamma_{n,t}^{\theta min}] \qquad (3e)$$

$$\theta_{n,t} = 0 , \quad \forall t, n = 1 \qquad [\gamma_t^1] \qquad (3f)$$

Constraints (3e) set the allowable limits for the voltage angles in order provide system stability, and (3f) arbitrarily sets the voltage angle to zero for bus number 1 (because the expression (3d) depends only on the difference in voltage angles between the two ends of each line).

Moreover, constraint (3g) is added to ensure that the power purchased by the transmission operator is balanced by its sales quantity, at each time period.

$$\sum_{n=1}^{N} \sum_{l=1}^{L} A_{l,n}^{grid} f_{l,t} = 0 , \quad \forall t \qquad [\gamma_t^{grid}] \qquad (3g)$$

$$f_{l,t}, \theta_{n,t} : Free, \quad \forall l, \forall t, \forall n \qquad (3h)$$

### 2.4. The SW model

In this section, the SW model is presented to maximize the social welfare, while setting electric energy and reserve prices, demand quantities, and generation dispatch at each time period. This model would be executed by the market operator, in order to send dispatch orders to market participants. This optimization model respects all the constraints of consumers and market participants. In addition, two more constraints (4b), and (4c) are added to enforce reserve capacity requirements and market clearing constraints for each bus.

The objective function of the SW model (4) is the summation of the objective functions of each of the models (1)-(3). However, due to reserve constraints (4b) and energy market clearing constraints (4c), the total payment by consumers will be equal to the total revenue of generators and transmission network operator and therefore their associated terms are canceled in the objective function of the SW problem. The complete SW model (4) is as follows:

$$\min \sum_{i=1}^{I}\sum_{t=1}^{T}[\lambda_{it}g_{it} + \lambda_{i,t}^{RES}A_i^{RES}g_{i,t}^R + C_{it}^{SU}v_{it} + C_{it}^{SD}w_{it} + C_{it}^{RP}r_{it}] \\ - \sum_{n=1}^{N}\sum_{t=1}^{T}[a_{n,t}q_{n,t} - 0.5b_{n,t}q_{n,t}^2] \quad (4a)$$

s.t.

$(1d), (2b) - (2n), (2p), (2q), (3b) - (3h),$

$$\sum_{i=1}^{I} A_i^{RES} g_{it}^R = \varphi \sum_n q_{n,t} \quad , \forall t \qquad [p_t^R] \qquad (4b)$$

$$\sum_{i=1}^{I} A_{i,n}^{gen} g_{i,t} = q_{n,t} + \sum_{l=1}^{L} A_{l,n}^{grid} f_{l,t} \quad , \quad \forall n, \forall t \qquad [p_{n,t}] \qquad (4c)$$

In practice, the dual variable prices $p_t^R$ and $p_{n,t}$ are calculated in two steps: first calculate the optimal solution to the MIP (4); then solve the LP which is (4) but with all binary variables fixed at the optimal values from the MIP, giving dual variables for all constraints, including for (4b) and (4c). In the next section, a near equilibrium unit commitment model is developed, where the approach suggested in Fuller and Çelebi (2017) is utilized to accommodate all realistic constraints of an electric system.

### 3. Developing the MTOC-MC Model

This section presents an extension to the MCP method which *approximates* opportunity-cost-minimizing prices, allowing for price-responsive demand, with quantities that are not obtained from the SW model of the previous section. The exact MTOC problem minimizes the total opportunity cost: the extra profit that market participants (generators and the transmission operator) could have had if they could just respond to prices. However, the exact MTOC problem cannot be solved directly; the approximation minimizes complementarities and is therefore called the MTOC-MC model. Here the opportunity cost is the difference between the profit when the instructions of the market operator are followed and when the market participants can freely make their own decision based on the market prices. We assume that consumers do respond optimally to prices, i.e., they have zero opportunity cost.

The following steps summarize the formulation of the MTOC-MC model:

1. For the profit maximization model of generation units, relax the binary variables to positive continuous variables and add constraints, $0 \leq z_{it} \leq 1$ to the model.

2. Develop the KKT conditions for consumers and all market participants, and combine them together with constraints that link participants together, i.e., the energy market clearing constraints and the reserve requirement constraints.
3. Replace the constraints $0 \leq z_{it} \leq 1$ with the binary constraints on $z_{it}$.
4. Replace the KKT conditions of consumers with linear expressions and new binary variables in the manner of Fortuny-Amat and McCarl (1981), to ensure that consumers respond optimally to prices.
5. Remove the complementarity conditions of market participants and set the objective function of MTOC-MC as minimizing the sum of their complementarity expressions. The remaining constraints are included as constraints in the MTOC-MC model.

In the following sections, the KKT conditions of market participants are developed to obtain the MTOC-MC model.

### 3.1. General form of MTOC-MC

In this section, a general form of MTOC-MC when firms' models are MILP is presented. For this purpose, some new notations utilizing Fuller and Çelebi (2017) are defined by Table 1 to explain the extension of the their model and the contribution of ours. We present this general form of MTOC-MC because the details of the original model worked out in this form in Fuller and Çelebi (2017), and also the inclusion of reserve capacities in our extended version of the model requires slight modification of the formulation presented in the original one.

Table 1. Notation for general form of MTOC-MC

| | |
|---|---|
| $f$ | Index of firms, i.e., generators and transmission operators |
| $A_f, B_f, D_f$ and $E_f$ | Matrix of coefficients |
| $x_f$ | Vector of continuous activities controlled by $f$ |
| $z_f$ | Vector of binary activities controlled by $f$ |
| $R$ | Matrix to relate reserve requirements to demands |
| $p$ and $p^R$ | Vector of energy and reserve prices |
| $q(\cdot)$ | Vector of demands as a function of energy and reserve prices |
| $c_f, d_f$, and $b_f$ | Vectors of coefficients |
| $\alpha_f$ | Vector of dual variables |
| $\alpha_f^z$ | Vector of dual variables for binary relaxation constraint |

Energy market clearing and system reserve constraints can be written as

$$\sum_f A_f x_f = q(p + R^T p^R) \qquad [p] \qquad (5a)$$

$$\sum_f B_f x_f = Rq(p + R^T p^R) \qquad [p^R] \tag{5b}$$

Constraints (5a) and (5b) link together the actions of the firms and the consumers. Fuller and Çelebi (2017) discussed only the linking constraints (5a), i.e., with the demand vector appearing without any matrix of coefficients. Our inclusion of (5b) is therefore a slight generalization; a quick review of the theory in Fuller and Çelebi (2017) confirms that the proofs of Theorems go through, provided the linking constraints are equality constraints. A general form of optimization problem for participating firms can be written as follows:

$$\min_{x_f, z_f} \left(c_f^T - p^T A_f - p^{R^T} B_f\right)x_f + d_f^T z_f \tag{5c}$$

s.t.

$$E_f x_f + D_f y_f \leq b_f \quad \forall f, \quad [\alpha_f] \tag{5d}$$

$$x_f \geq 0, z_f \in \{0,1\}^{\dim(z_f)}, \forall f$$

For a transmission network operator, vectors $c_f$ and $d_f$ are zero ($c_f = d_f = 0$) and $z_f$ is a null vector, i.e., $\dim(z_f) = 0$. In the general form, any equality constraint desired by the modeler can be represented by two inequalities and any free variable can be replaced by a difference between two non-negative variables.

By following the four-step procedure above, the MTOC-MC includes all primal constraints of the firms' optimization models, together with all of the dual constraints of the binary-relaxed firms' models, and also the energy market clearing and system reserve requirement constraints. In the formulation of the sum of complementarities objective function, there are many cancellations. Defining $\alpha_f^z$, as the dual variable vector of the relaxation constraint, $z_f \leq 1$ ($\dim(\alpha_f^z) = \dim(z_f)$), a general form of the MTOC-MC model can be represented as follows:

$$\min_{x_f, z_f, p, p^R, \alpha_f, \alpha_f^z} \sum_f \left\{ b_f^T \alpha_f + 1^T \alpha_f^z + \left(c_f^T - p^T A_f - p^{R^T} B_f\right)x_f + d_f^T z_f \right\} \tag{5e}$$

s.t.

$$b_f - E_f x_f - D_f z_f \geq 0, \quad \forall f \tag{5f}$$

$$c_f - A_f^T p - B_f^T p^R + E_f^T \alpha_f \geq 0, \quad \forall f \tag{5g}$$

$$d_f + D_f^T \alpha_f + \alpha_f^z \geq 0, \quad \forall f \tag{5h}$$

$$q(p + R^T p^R) - \sum_f A_f x_f = 0 \tag{5i}$$

$$Rq(p + R^T p^R) - \sum_f B_f x_f = 0 \tag{5j}$$

$$\alpha_f \geq 0, \alpha_f^z \geq 0, x_f \geq 0, z_f \in \{0,1\}^{\dim(z_f)}, \quad \forall f \tag{5k}$$

The MTOC-MC model is a mixed integer nonlinear program because of nonlinear terms, $\left(-p^T A_f - p^{R^T} B_f\right) x_f$ and $q(p + R^T p^R)$ if this function is nonlinear. We show in section 3.2 that if we specify the function $q(p + R^T p^R)$ as the optimal response of consumers to the electricity and reserve prices, using the consumers' model of section 2.1, then we can formulate the MTOC-MC model as a mixed integer quadratic convex program (MIQCP) having linear constraints. In particular, the nonlinear objective terms $\left(-p^T A_f - p^{R^T} B_f\right) x_f$ are replaced by convex linear quadratic expressions. This formulation allows us to use the reliable and efficient MIQCP solver in CPLEX.

### 3.2. Optimal response of consumers and MIQCP form of MTOC-MC

The KKT conditions for the convex optimization model of consumers, (1c)-(1d) are:

$$0 \leq q_{dn,t} - Q_{dn,t}^{min} \perp \beta_{dn,t} \geq 0, \quad \forall dn, t \tag{6a}$$

$$p_{dn,t} + \varphi p_t^R - a_{dn,t} + b_{dn,t} q_{dn,t} - \beta_{dn,t} = 0, \quad \forall dn, t \tag{6b}$$

Where, $dn$ is the index for demand buses. Using the technique of Fortuny-Amat & McCarl (1981), the consumers' KKT conditions (6a) can be represented equivalently by the following linear constraints ($u_{dn,t}$ is a new binary variable, $M$ is a big number, and the dual variable $\beta_{dn,t}$ can be eliminated by using the equality of (6b).

$$q_{dn,t} \geq Q_{dn,t}^{min}, \quad \forall dn, t \tag{6c}$$
$$p_{dn,t} + \varphi p_t^R - a_{dn,t} + b_{dn,t} q_{dn,t} \geq 0, \quad \forall dn, t \tag{6d}$$
$$p_{dn,t} + \varphi p_t^R - a_{dn,t} + b_{dn,t} q_{dn,t} \leq M u_{dn,t}, \quad \forall dn, t \tag{6e}$$
$$q_{dn,t} - Q_{dn,t}^{min} \leq M(1 - u_{dn,t}), \quad \forall dn, t \tag{6f}$$

The nonlinear terms of the MTOC-MC objective function (5e) can be converted to expressions in the variables of the consumers' model using the reserve and electricity market clearing constraints (4b)-(4c), multiplied by their prices $p_t^R$ and $p_{n,t}$ respectively.

$$\left(-p^T A_f - p^{R^T} B_f\right) x_f = -\sum_{i,t,n} A_i^{gen} g_{i,t} p_{n,t} - \sum_{i,t} A_i^{Res} g_{i,t}^R p_t^R + \sum_{l,t,n} A_{l,n}^{grid} f_{l,t} p_{n,t}$$

$$= -\sum_{dn,t}(p_{n,t} + \varphi p_t^R) q_{n,t}, \text{ using (4b), (4c)}$$

$$= \sum_{dn,t}\{-a_{dn,t} q_{dn,t} + b_{nd,t} q_{nd,t}^2 - \beta_{dn,t} q_{dn,t}\}, \text{ using (6b)}$$

$$= \sum_{dn,t}\{-a_{dn,t} q_{dn,t} + b_{nd,t} q_{nd,t}^2 - \beta_{dn,t} Q_{dn,t}^{min}\}, \text{ using (6a)}$$

$$= \sum_{dn,t}\{-a_{dn,t}q_{dn,t} + b_{nd,t}q_{nd,t}^2 - (p_{dn,t} + \varphi p_t^R - a_{dn,t} + b_{dn,t}q_{dn,t})Q_{dn,t}^{min}\}, \text{ using (6a)}.$$

The term $a_{dn,t}Q_{dn,t}^{min}$ is constant and can be dropped from the objective function. The only nonlinear term is, $b_{nd,t}q_{nd,t}^2$, which is convex and so the MTOC-MC converts to a mixed integer quadratic convex problem (MIQCP) and can be solved efficiently by CPLEX.

### 3.3. The KKT constraints of generation units and transmission network operator

The firms' primal constraints (5f) in the general MTOC-MC model correspond to generators' and transmission operator's constraints (2b) to (2n), (2p), (2q) and (3b) to (3h). The firms' dual constraints (5g), (5h) and the non-negativity conditions on dual variables in (5k) correspond to the following constraints which are the derivative and non-negativity conditions of the KKT conditions for the generators and the transmission operator. (The corresponding primal variables are shown in brackets to the right of the constraints.)

$$-\lambda_{it} + \sum_n A_{i,n}^{gen} p_t - \alpha_{it}^{RP1} + \alpha_{it+1}^{RP1} - \alpha_{it+1}^{RP2} + \alpha_{it}^{RP2} - \alpha_{it}^{RP3} + \alpha_{it+1}^{RP3} - \alpha_{it+1}^{RP4} + \alpha_{it}^{RP4}$$
$$+ \alpha_{it}^{min} - \alpha_{it}^{max} \leq 0 \,, \forall i,t \qquad [g_{it}] \quad (7a)$$

$$-C_{it}^{SU} + \alpha_{it}^{RP3}(SU_i - RU_i) - \alpha_{it}^{UT}U_i^{min} + \alpha_{it}^{OP} - \sigma_{i,t} \leq 0 \,, \forall i,t \qquad [v_{it}] \quad (7b)$$

$$-C_{it}^{SD} + \alpha_{it}^{RP4}(SD_i - RD_i) - \alpha_{it}^{DT}D_i^{min} - \alpha_{it}^{OP} \leq 0 \,, \forall i,t \qquad [w_{it}] \quad (7c)$$

$$\alpha_{it}^{RP3}RU_i + \alpha_{it+1}^{RP4}RD_i - \alpha_{it}^{min}G_i^{min} + \alpha_{it}^{max}G_i^{max} - \sum_{k=\max(t-D_i^{min}+1,\,1)}^{t} \alpha_{ik}^{DT}$$
$$+ \sum_{k=\max(t-U_i^{min}+1,\,1)}^{t} \alpha_{ik}^{UT} - \alpha_{it}^{OP} + \alpha_{it+1}^{OP} - \alpha_{it}^{Z} - \sigma_{i,t+1} \leq 0 \,, \forall i \qquad [z_{it}] \quad (7d)$$
$$\notin \{I_{ON} \cup I_{OFF}\}, \forall t$$

$$\alpha_{it}^{RP3}RU_i + \alpha_{it+1}^{RP4}RD_i - \alpha_{it}^{min}G_i^{min} + \alpha_{it}^{max}G_i^{max} - \sum_{k=\max(t-D_i^{min}+1,\,1)}^{t} \alpha_{ik}^{DT}$$
$$+ \sum_{k=\max(t-U_i^{min}+1,\,1)}^{t} \alpha_{ik}^{UT} - \alpha_{it}^{OP} + \alpha_{it+1}^{OP} - \alpha_{it}^{Z} - \sigma_{i,t+1} + \alpha_{it}^{ON} \qquad [z_{it}] \quad (7e)$$
$$\leq 0 \,, \forall i \in I_{ON}, \forall t$$

$$\alpha_{it}^{RP3}RU_i + \alpha_{it+1}^{RP4}RD_i - \alpha_{it}^{min}G_i^{min} + \alpha_{it}^{max}G_i^{max} - \sum_{k=\max(t-D_i^{min}+1,\,1)}^{t} \alpha_{ik}^{DT}$$
$$+ \sum_{k=\max(t-U_i^{min}+1,\,1)}^{t} \alpha_{ik}^{UT} - \alpha_{it}^{OP} + \alpha_{it+1}^{OP} - \alpha_{it}^{Z} - \sigma_{i,t+1} + \alpha_{it}^{OFF} \leq 0 \,, \forall i \qquad [z_{it}] \quad (7f)$$
$$\in I_{OFF}, \forall t$$

$$\alpha_{it}^{RP1} + \alpha_{it}^{RP2} - C_{it}^{RP} \leq 0 \,, \qquad \forall i,t \qquad [r_{it}] \quad (7g)$$

$$A_i^{RES}p_t^R - A_i^{RES}\alpha_{i,t}^{max} - A_i^{RES}\lambda_{i,t}^{RES} \leq 0, \qquad \forall i,t \qquad [g_{it}^R] \quad (7h)$$

$$\sum_{n=1}^{N} A_{l,n}^{grid} p_{n,t} - \sum_{n=1}^{N} A_{l,n}^{grid} \gamma_t^{grid} + \gamma_{l,t}^{max} - \gamma_{l,t}^{min} + \gamma_{l,t}^{phyc} = 0, \qquad \forall l, \forall t \qquad [f_{lt}] \quad (7i)$$

$$\sum_{l} B_l A_{l,n}^{grid} \gamma_{l,t}^{phyc} - \gamma_{n,t}^{\theta max} + \gamma_{n,t}^{\theta min} = 0 \qquad [\theta_{nt}] \quad (7j)$$

$$r_{it}, g_{it}, v_{it}, w_{it}, \alpha_{it}^{RP1}, \alpha_{it}^{RP2}, \alpha_{it}^{RP3}, \alpha_{it}^{RP4}, \alpha_{it}^{min}, \alpha_{it}^{max}, \alpha_{it}^{DT}, \alpha_{it}^{UT}, \alpha_{it}^{z}, \sigma_{i,t} \geq 0, \forall i, t \quad (7k)$$

$$\alpha_{it}^{OP}, \alpha_{it}^{ON}, \alpha_{it}^{OFF}: \text{Free} \quad (7l)$$

$$z_{it} = \{0,1\}, \forall i, t \quad (7m)$$

$$f_{l,t}, \theta_{n,t}, \gamma_t^{grid}, \gamma_{l,t}^{phyc}: Free, \gamma_{l,t}^{max}, \gamma_{l,t}^{min}, \gamma_{n,t}^{\theta max}, \gamma_{n,t}^{\theta min} \geq 0, \forall l, \forall t, \forall n \quad (7n)$$

### 3.4. The MTOC-MC model

The objective function of the MTOC-MC is formed below as (8a), considering the general form (5e) applied to the generators' and transmission operators' models, together with the results of subsection 3.2 which replace nonlinear terms by convex linear-quadratic expressions.

$$\begin{aligned}
\min \sum_i \sum_t & \left[ \lambda_{it} g_{it} + \lambda_{i,t}^{RES} A_i^{RES} g_{i,t}^R + C_{it}^{SU} v_{it} + C_{it}^{SD} w_{it} + C_{it}^{RP} r_{it} + \alpha_{it}^z + \sigma_{i,t} \right] \\
& - \sum_{i \in I_{ON}} \sum_{t=1}^{t_i^{ON}} \alpha_{it}^{ON} \\
& + \sum_i \left[ \alpha_{i1}^{RP1} g_{i0} - \alpha_{i1}^{RP2} g_{i0} + \alpha_{i1}^{RP3} g_{i0} - \alpha_{i1}^{RP4} g_{i0} + \alpha_{i1}^{RP4} RD_i z_{i0} + \alpha_{i1}^{OP} z_{i0} \right. \\
& \left. + \sum_t D_i^{min} \alpha_{it}^{DT} \right] \\
& + \sum_n \sum_t \left[ -a_{n,t} q_{n,t} + b_{n,t} q_{n,t}^2 - (p_{n,t} + \varphi p_t^R) Q_{n,t}^{min} - Q_{n,t}^{min} b_{n,t} q_{n,t} \right] \\
& + \sum_{t=1}^T \sum_{l=1}^L \left[ F_l^{max} (\gamma_{l,t}^{max} + \gamma_{l,t}^{min}) + \pi (\gamma_{n,t}^{\theta max} + \gamma_{n,t}^{\theta min}) \right]
\end{aligned} \quad (8a)$$

Furthermore, the MTOC-MC model's constraints are determined by combining the constraints obtained from the KKT conditions of optimization problems of consumers and market participants, as well as energy and reserve market clearing constraints, as follows:

(2b) − (2n), (2p), (2q), (3b) − (3h), (4b), (4c), (6c) − (6f), (7a) − (7n)

Fuller and Çelebi (2017) proposed another calculation after solving the MTOC-MC model that considers the prices and demands obtained by MTOC-MC as given parameters and then solves the minimum cost UC problem. This approach produced instructions for generators that reduced the total system cost and also reduced the total opportunity cost, compared to the MTOC-MC solution. However, for the example data set of this paper, this extra optimization calculation makes no improvement and so is not presented in this paper.

## 4. Computational results

In this section, we examine the performance of SW and MTOC-MC models by applying them on the modified IEEE RTS 24-bus test system (Ordoudis et al., 2016). To conduct this study, reserve requirement $\varphi$ is assumed to be 5% of demand, and reserve capacity cost, $\lambda_{it}^{Res}$ is 40% of the operational cost, $\lambda_{it}$. Additionally, the startup and shutdown rates ($SU_i$ and $SD_i$) are considered to be equal to the ramp up and ramp down rates ($RU_i$ and $RD_i$), respectively. Also, the shutdown cost, $C_{it}^{SD}$ is assumed to be equal to 20% of startup cost, $C_{it}^{SU}$. The line susceptance, $B_l$ is calculated as the inverse of its reactance (Gabriel et al, 2013b).

In Ordoudis et al. (2016) the demand is assumed to be fixed at each node and time period. However, in this paper we assume that part of the demand is flexible, meaning that consumers can adjust their consumption based on prices at each time period. If the given demand in Ordoudis et al. (2016) is shown by $\bar{Q}_{n,t}$, the fixed part of the demand is $Q_{n,t}^{min}$ and the rest of it $\bar{Q}_{n,t} - Q_{n,t}^{min}$ is flexible. The value of $Q_{n,t}^{min}$ is approximated by assuming that $\frac{\bar{Q}_{n,t} - Q_{n,t}^{min}}{\bar{Q}_{n,t}} \leq 0.15$ or at least 85% of the given demand in Ordoudis et al. (2016) is fixed and the rest is flexible.

In order to choose the values of $a_{n,t}$ and $b_{n,t}$ in (1a) lower and higher electricity price of $P_{n,t}^l$ and $P_{n,t}^h$ are assumed at each node and time period. The two points $(Q_{n,t}^{min}, P_{n,t}^h)$ and $(\bar{Q}_{n,t}, P_{n,t}^l)$ are assumed to be on the line, shown by (1a), thus determining $a_{n,t}$ and $b_{n,t}$.

To perform sensitivity analysis on the flexible part of demand, equations (1d) and (6c) are changed as follows:

$$q_{n,t} \geq hQ_{n,t}^{min}, \quad \forall n, t \tag{9a}$$

Where, $h$ is a scalar that can take a non-negative value. The symbol $Q_{n,t}^{min}$ is also replaced by $hQ_{n,t}^{min}$, in (6f) and (8a). The value of $h = 0$ reflects a market such that 100% of the demand is responsive to price. The default model is when $h = 1$ and when $h > 1$, a smaller portion of demand is price-responsive demand than the default demand model. We assume that parameter $h$ can take a value in the following range:

$$0 \leq h < \min_{\forall n,t} \left\{ \frac{\bar{Q}_{n,t}}{Q_{n,t}^{min}} \right\}$$

The proposed optimization models are formulated in the GAMS 24.7.4 environment, and the CPLEX 12.6.3.0 solver is used to solve the proposed LPs and MIPs, on a server, with a 64-bit operating system, with 39 GB of RAM, and dual core (8 processors), 2.61 GHz CPU. The size of MTOC-MC and SW models, with respect to the number of binary and continuous variables, constraints and their execution time is presented in Table 2 for the $h = 1$ case. As shown, the larger size of the MTOC-MC model due to inclusion of dual variables, extra binary variables and extra constraints, has an impact on execution time, but the time is still very short for practical use. In order to have accurate comparison results, we set the "optcr" equal to zero in the code to have the optimal solution for both models.

Table 2. Comparison of models' size and running time

| Description | MTOC-MC | SW |
|---|---|---|
| Number of binary variables | 696 | 288 |
| Number continuous variables | 11281 | 3457 |
| Number of constraints | 11809 | 7496 |
| Solution run time (seconds) | 6.75 | 1.44 |

### 4.1. Numerical analysis ($h = 1$)

In this section, the results of running the MTOC-MC and SW models are compared for the case of $h = 1$ in (9a), (6f) and (8a). The dispatch instructions of the two models are not the same, and hence the start-up and shutdown costs, total generation cost, revenue, and profit of generators are different in each model.

As shown in Table 3 the total profit of generation units is larger in the solution of the MTOC-MC model compared to the solution of SW. The generation units, in total, gain more revenue and incur less cost by the solution of MTOC-MC, and therefore create higher total profit. Almost all generators (except GEN10) have a larger profit in MTOC-MC than in SW. If the system operator pays a make-whole payment to compensate for negative profit (e.g., GEN3 in Table 3), it is observed that MTOC-MC requires smaller value of make-whole payment than SW; furthermore, the make-whole payments are a negligible addition to consumers' payments.

Table 3. Financial impact of MTOC-MC and SW models on generation units ($h = 1$)

| GEN# | MTOC-MC | | | SW | | |
|---|---|---|---|---|---|---|
| | Total cost | Total Revenue | Total Profit | Total cost | Total Revenue | Total Profit |
| GEN1 | 38,008.5 | 41,439.8 | 3,431.2 | 38,026.0 | 41,338.3 | 3,312.3 |
| GEN2 | 20,396.8 | 22,669.8 | 2,273.0 | 20,404.7 | 22,570.6 | 2,165.8 |
| GEN3 | 104,046.4 | 103,956.9 | -89.5 | 107,616.0 | 106,919.1 | -696.9 |
| GEN4 | 0.0 | 0.0 | 0.0 | 0.0 | 0.0 | 0.0 |
| GEN5 | 0.0 | 0.0 | 0.0 | 0.0 | 0.0 | 0.0 |
| GEN6 | 32,142.8 | 59,203.9 | 27,061.1 | 32,142.8 | 58,909.0 | 26,766.2 |
| GEN7 | 39,134.4 | 69,543.5 | 30,409.1 | 39,134.4 | 69,246.4 | 30,112.0 |
| GEN8 | 55,986.0 | 177,262.3 | 121,276.3 | 55,384.2 | 175,469.8 | 120,085.6 |
| GEN9 | 52,217.9 | 178,652.7 | 126,434.8 | 52,512.0 | 178,212.4 | 125,700.4 |
| GEN10 | 0.0 | 134,022.4 | 134,022.4 | 0.0 | 133,449.6 | 133,449.6 |
| GEN11 | 78,268.8 | 140,037.3 | 61,768.5 | 78,268.8 | 139,434.4 | 61,165.6 |
| GEN12 | 88,867.4 | 152,138.1 | 63,270.6 | 88,867.4 | 151,457.4 | 62,590.0 |
| SUM: | 509,069.1 | 1,078,926.7 | 569,857.5 | 512,356.3 | 1,077,006.9 | 564,650.6 |

Table 4 summarizes the financial performance of consumers and market participants. As shown in Table 4, the MTOC-MC solution is better than the SW's, in terms of total generation units' cost and profit. Generation units gain 0.92% more profit by the solution of the MTOC-MC model compared to the SW's, while consumers only pay 0.18% more in the solution of the MTOC-MC model. MTOC-MC or SW results are the same for the profit of transmission network operators. Although the consumers' surplus is slightly less in MTOC-MC (-0.51%) compared to SW, the total social welfare is almost the same for both models. Therefore, using the results of this test system, it can be concluded that the MTOC-MC works slightly in favor of generation units by improving their costs and net profit.

Table 4. The summary of financial performance of MTOC and SW models

| Description | MTOC-MC | SW | Change in MTOC-MC over SW (%) |
|---|---|---|---|
| Payment by consumers | 1,080,613.9 | 1,078,694.2 | 0.18% |
| Consumers' surplus | 1,036,316.9 | 1,041,596.9 | -0.51% |
| Total generation cost | 509,069.1 | 512,356.3 | -0.64% |
| Total revenue of generation units | 1,078,926.7 | 1,077,006.9 | 0.18% |
| Total profit of generation units | 569,857.5 | 564,650.6 | 0.92% |
| Profit of transmission network operator | 1,687.3 | 1,687.3 | 0.00% |
| Total social welfare | 1,607,861.7 | 1,607,934.7 | -0.0045% |

Figures 2, 3, and 4 illustrate the total demand, weighted average nodal energy prices (total revenue from consumers divided by total demand), and reserve capacity prices, respectively, over each time period. As shown by Figure 2, the total demand determined by the two models closely matches each other, with slight differences mostly in peak demand periods, when MTOC-MC demand is smaller than in the SW model.

Figure 3 illustrates the average nodal energy prices for both MTOC-MC and SW models over each of the periods. As shown by this figure, the average energy prices determined by both models are the same in most base, and medium periods, while during the peak periods the prices derived from the MTOC-MC model are slightly higher than the SW prices.

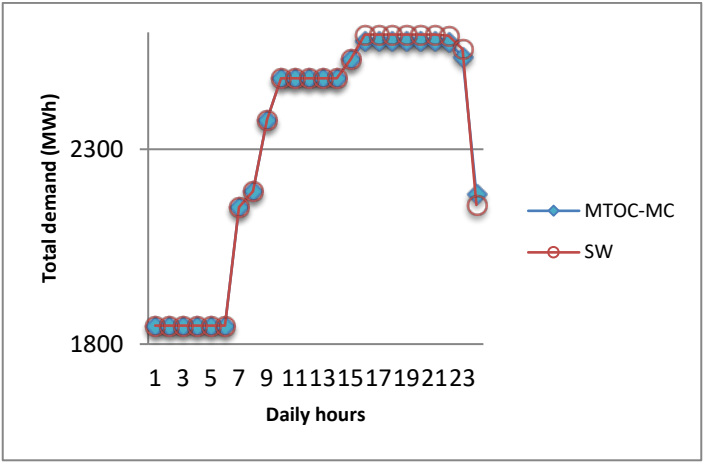

Figure 2. Total demand over periods

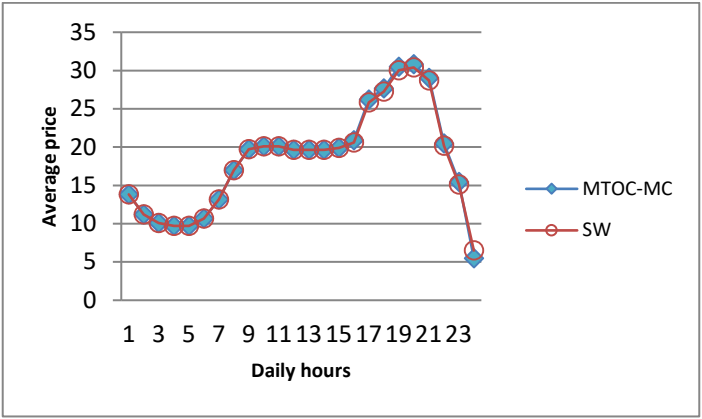

Figure 3. Average nodal energy prices over periods

Figure 4 illustrates the reserve prices determined by both models. Again, prices are the same for the two models in most base and medium demand periods, but the MTOC-MC model has slightly higher reserve prices for peak periods. Another important criterion to compare the performance of the models is total opportunity cost of the system, as presented in Table 5. Here the total opportunity cost is defined as the difference between the total profit that the market participants could earn following the direction of the market operator, and the profit that they could earn if each firm followed its own decision based only on the prices announced by the system operator.

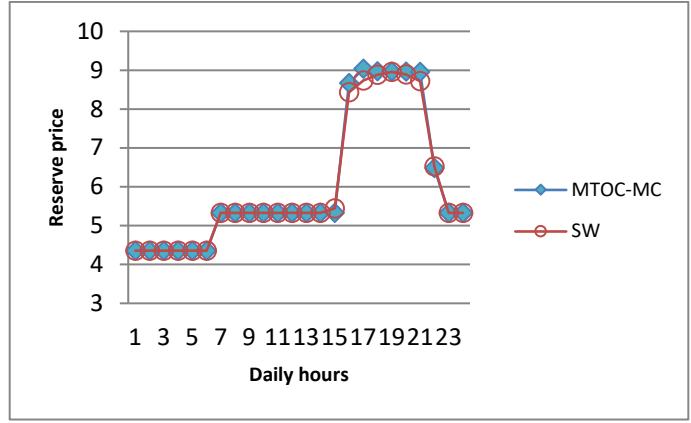

Figure 4. The reserve capacity price over periods

Consumers are also included in Table 5, with consumers' surplus in place of profit to determine opportunity costs which, by model design, are necessarily zero.

Table 5. Total opportunity cost

| Participants | MTOC-MC | SW |
|---|---|---|
| Generation units | 118.4 | 1,123.1 |
| Transmission network operator | 0.0 | 0.0 |
| Consumers | 0.0 | 0.0 |
| SUM: | 118.4 | 1,123.1 |

As shown in Table 5, the total opportunity cost of the system decreases 89.4% by the solution of MTOC-MC compared to SW. While the generators' opportunity cost decreases significantly, there is no change in the opportunity cost of the transmission network operator and consumers, as they are zero in both models. If the system operator were to pay generators all of their opportunity costs (not just make-whole payments), then for both models there is a very small addition to consumers' payment with the MTOC-MC payment being extremely small compared to total payments by consumers.

In summary, the MTOC-MC model comes closer to equilibrium than the SW model, as indicated by total opportunity cost and by the more restricted make-whole payment measures. This improvement comes at the cost of a very small decrease in social welfare MTOC-MC compared to SW. Consumers face slightly higher prices and have lower demands in the peak periods, leading to small increase in consumer payments and a small decrease in consumers' surplus.

## 4.2. The impact of demand price responsiveness

In this subsection, different values of the parameter $h$ (0.00, 0.25, 0.50, 0.75, 1, and 1.2) in (9a) are considered to investigate the impact of demand flexibility. Among the tested values, $h = 0$ represents the situation, where there is no fixed demand, and $h = 1.2$ represents the situation when a larger part of demand is fixed. The comparison between results from the SW and MTOC-MC model with different values of $h$ is shown in Figure 5.

In both models, as the value of $h$ increases, the total profit of Generation Units (GENs) increases, while the profit of Transmission Operator (TO) and Consumers Surplus (CS) decreases. In other words, generation units gain more profit in the market, when demand is less responsive to price. When the amount of responsive demand increases, the profit of GENs declines, and TO gains more profit and the CS value increases. The solution of the MTOC-MC model offers higher profit for GENs than the SW one, when the value of $h$ is less than or equal to one. On the other hand, the SW solution provides more profit for GENs, when the value of $h$ is greater than one.

As demand becomes less responsive to price (value of $h$ increases), the total demand and prices increase in both models. The MTOC-MC model offers slightly higher prices and lower demand compared to the SW model, when $h$ values are less than or equal to one. In other words, in the cases when $h \leq 1$, MTOC-MC gives higher electricity prices and lower demand quantities than the SW model. On the other hand, when demand is fixed or a very small portion of demand is price responsive ($h > 1$), SW gives higher prices and lower demand quantities than MTOC-MC.

Figure 6 illustrates the comparison of the total opportunity cost and total social welfare resulting from the solutions of the SW and MTOC-MC models with different values of $h$. As expected, the total opportunity cost is less in the MTOC-MC than in the SW solutions. The solutions of both models show an almost steady total opportunity cost, for all values of $h < 1$. However, the difference between the total opportunity cost obtained from both models increases significantly, for values of $h > 1$, when demand is less flexible. For the case $h = 1.2$, the total opportunity cost obtained by the solution of SW is more than 4 times as much as the one obtained by the MTOC-MC model. In other words, MTOC-MC performs better than SW with respect to the total opportunity cost measure, especially in markets with mostly fixed demand.

As shown in Figure 6, the total social welfare decreases as $h$ increases for both models. In other words, the total social welfare increases as flexibility of demand increases. In general, the SW model gives higher social welfare than MTOC-MC. In the markets with high flexible demand, the difference between two models becomes very small. However, for $h = 1.2$ which reflects a market with mostly fixed demand, the difference of total social welfare becomes larger. In summary, we conclude that changing the amount of price-responsive demand has little impact on the difference between the solutions of the MTOC-MC and SW models, except when demand is mostly fixed.

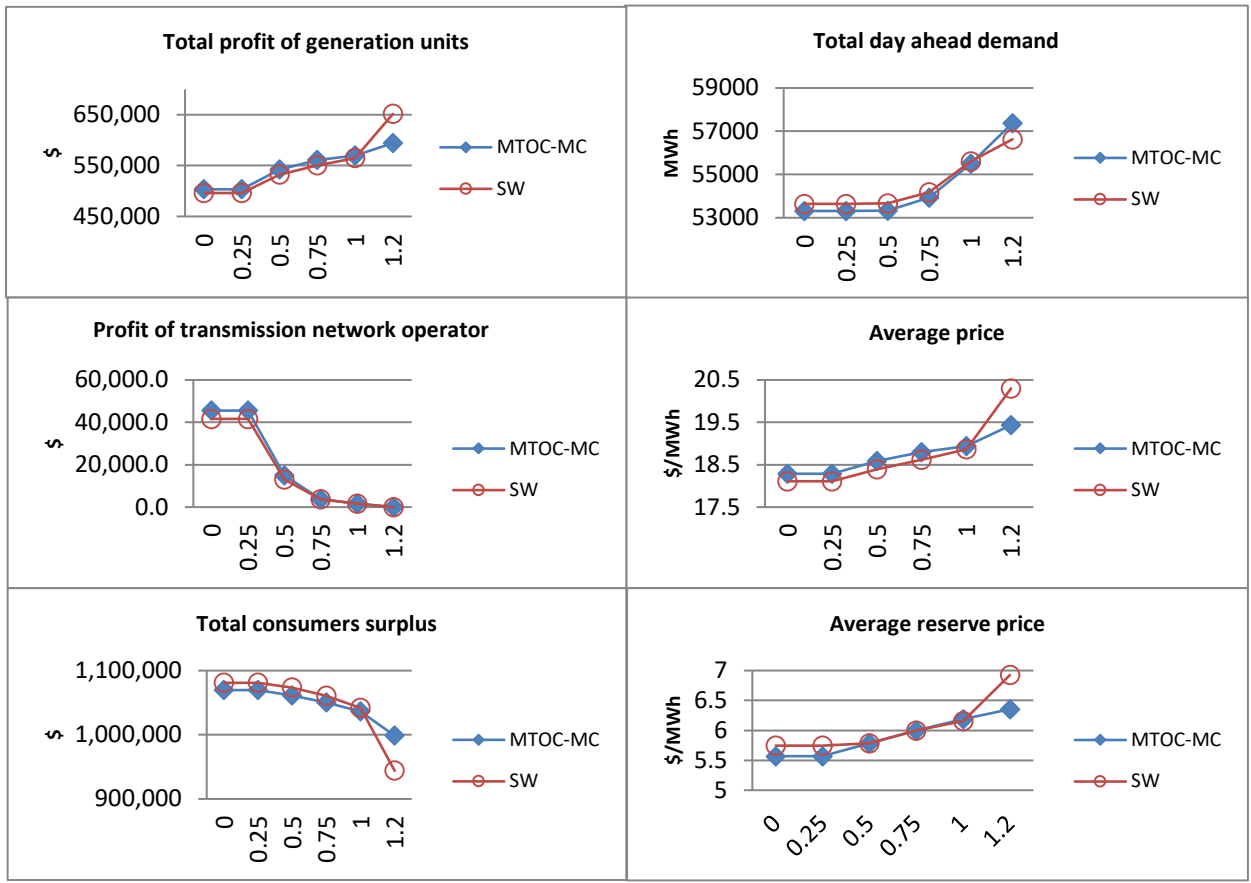

Figure 5. Comparative results on different values of $h$

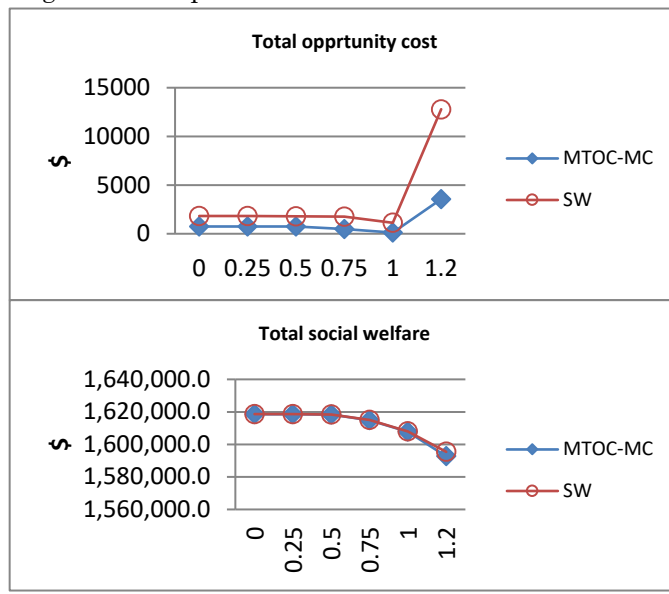

Figure 6. The total opportunity cost and total social welfare for different values of $h$

## The Impact of maximum line flow capacity

In this subsection, the impact of maximum line flow capacity $F_l^{max}$ on the solution of both models is investigated. In order to perform such sensitivity analysis, we vary $F_l^{max}$ for all lines simultaneously, from 60% to 100% of their initial given values in the original dataset. The lines are more congested when maximum capacity is at 60% of $F_l^{max}$, and are less congested when the lines are at 100% of their capacity. Figure 7 summarizes the results obtained from the solution of both models over different values of $F_l^{max}$.

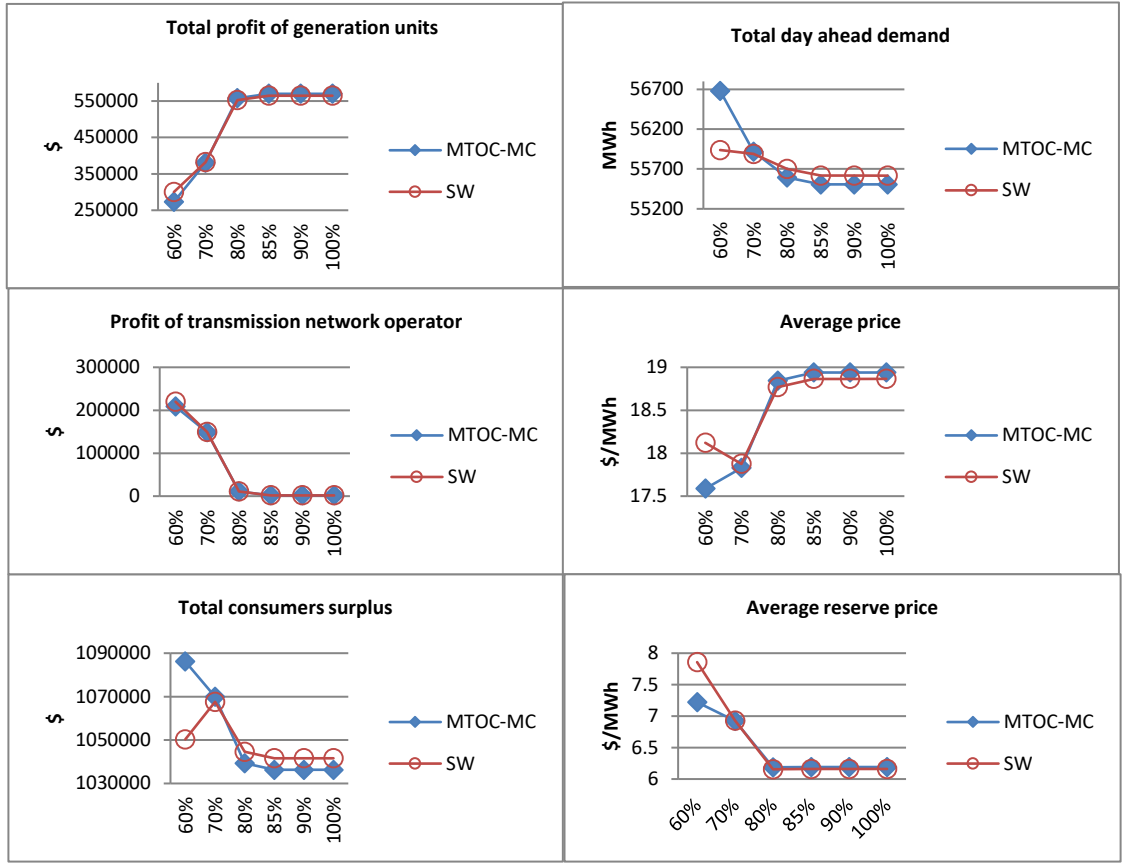

Figure 7. Comparative results of models on different percentages of $F_l^{max}$

As shown, the total profit of GENs grows as the maximum capacity of lines $F_l^{max}$ increases. However, at about 80% of line capacity the profit change is small, as the prices become steadier after that threshold, due to less congestion in transmission lines. In contrast, the profit of TO and CS decreases in the solutions of both models as the line capacity becomes more available. The solution of both models are very close to each other for both total generation and transmission network profit. However, the difference becomes more significant considering the value of CS. The solution obtained by SW generates higher values of CS for cases when line capacity is more available (≥70% $F_l^{max}$), while MTOC-MC provides better solution for CS for cases when line capacity is tighter ( <70% $F_l^{max}$).

Average prices (total daily demand) increase (decrease) as the max available line flows increase, except when $F_l^{max}$ is at 60% for the solution of the SW model. MTOC-MC offers slightly higher prices (lower total demand) than SW when line capacities are more available (>70%$F_l^{max}$). The average reserve price decreases as $F_l^{max}$ becomes larger and SW offers higher reserve price for $F_l^{max}$ less than 70% of the original value.

In all instances, i.e., more or less congested systems, the total opportunity cost is higher for the solution of the SW model, as expected. As illustrated in Figure 8, this difference is more significant in more congested systems, when line capacities are less than or equal to 70% of the original capacity. The total social welfare as illustrated in Figure 8, increases when increasing $F_l^{max}$. However, beyond a threshold of 85%, it remains constant. The difference between SW and MTOC-MC is not considerable although SW offers a bit higher social welfare in all cases.

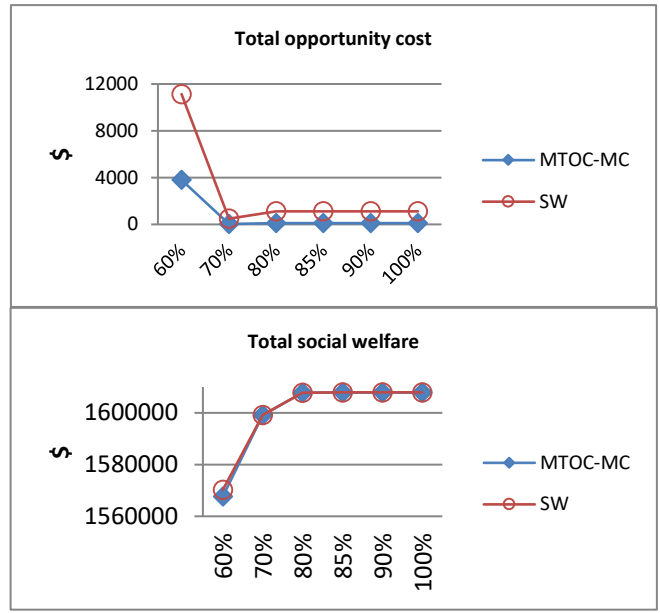

Figure 8. The total opportunity cost and total social welfare of models on different values of %$F_l^{max}$

The two interesting observations from this sensitivity analysis are:

- The difference between MTOC-MC and SW is significant in highly congested networks.
- For most cases, consumers' surplus is smaller in the MTOC-MC solution than in the SW solution, but this is reversed in the very congested cases.

### 4.4. The impact of maximum generation capacity

In this section, we study the impact of the available generation capacity or the value of $G_i^{max}$ on the solution of the MTOC-MC and SW models. We perform a sensitivity analysis, by changing the value of $G_i^{max}$ from 70% to 130% of the given initial value in the dataset. Figure 9 summarizes the comparative results, obtained from the sensitivity analysis.

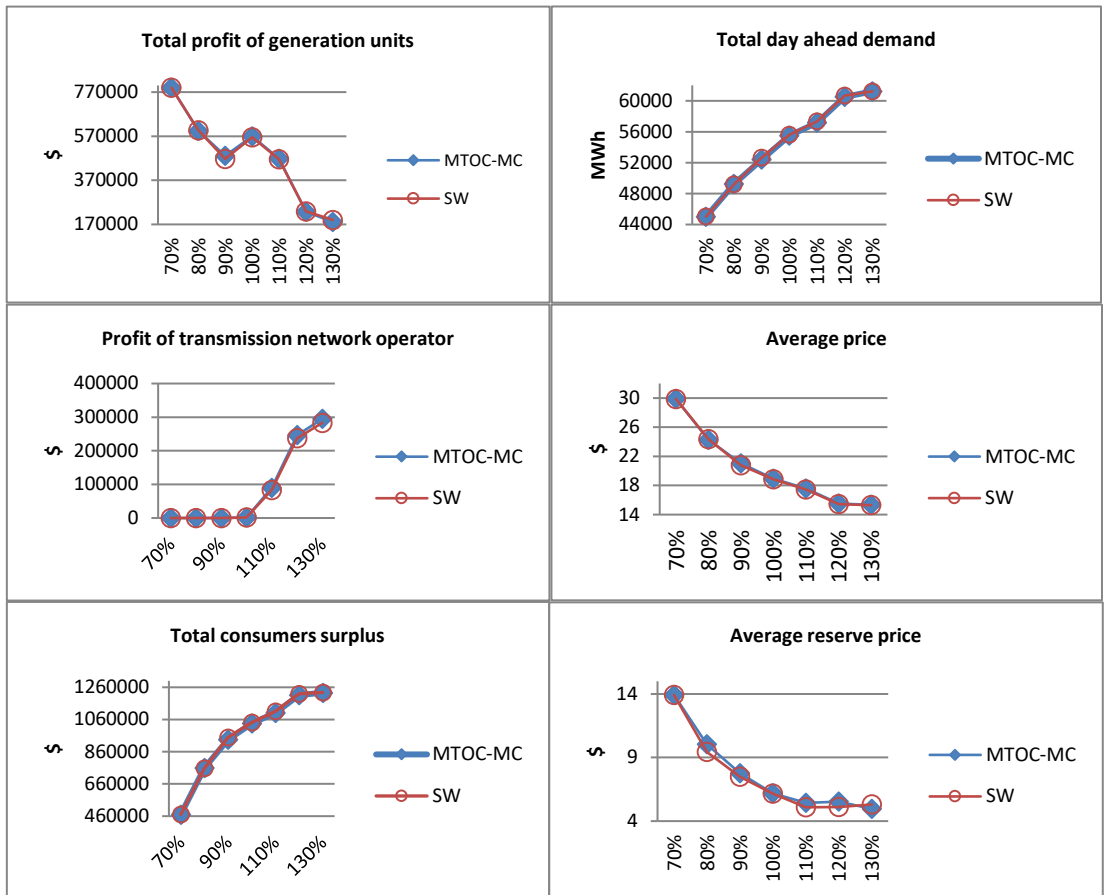

Figure 9. Comparative results on different values of %$G_i^{max}$

As shown, the solution of the two models matches each other closely, and there is no significant difference between MTOC-MC and SW results. The total profit of GENs decreases as the max available capacity increases. In addition, the profit of TO is almost zero up the 100% of max allowed capacity, and then starts increasing as the $G_i^{max}$ value becomes higher than 100%. This shows that the increase of available capacity of GENs is in favor of TO, not the GENs. Also, consumers gain more surpluses by increasing $G_i^{max}$ due to decrease of prices and rise of demand.

Figure 10 illustrates the total opportunity cost and total social welfare of both models over different values of maximum generation capacity. Although the difference between the solution of MTOC-MC and SW is not significant, the difference between the total opportunity cost obtained by the two models is considerable. The total opportunity cost increases at both models, when $G_i^{max}$ is over 100% of the initial given capacities in the dataset. In both models, the total social welfare increases when $G_i^{max}$ increases. There is no significant difference between the two models.

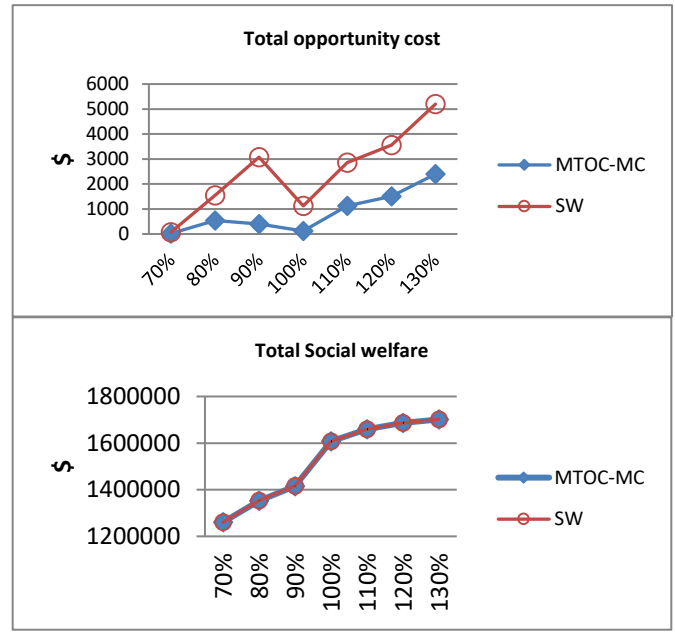

Figure 10. The total opportunity cost and total social welfare of models for different values of $G_i^{max}$

## 5. Directions for further research

Most real electric power systems are much larger than our 24 bus, 12 generator test system – real systems can have on the order of thousands of buses and a hundred or more generators. For our small test system, the MTOC-MC model took about five times longer to compute a solution than for the SW model. If the MTOC-MC model is to be considered for use in real systems, then further research may be needed into methods to speed up the computational time. A first step might be to aggregate the real transmission system into a "zonal" system, with a few fictitious buses representing many buses and lines within zones.

Our demand model is extremely simple, having a linear marginal value function of quantity, with a lower bound on quantity demanded, and each period's demand is independent of conditions in other periods. A more realistic, flexible demand could incorporate the notion that demand could be shifted by consumers, perhaps using ideas such as in Zoltowska (2016b). Such a demand could be partially dispatchable by the ISO which could choose the period, within limits, but the consumer would still respond optimally to prices in the MTOC_MC model, i.e., consumers would have zero opportunity costs.

For systems with a large fraction of renewable but intermittent power production such as wind and solar, the MTOC-MC model could be modified using some of the stochastic programming ideas presented by Abbaspourtorbati et al. (2017) and Zhan and Zheng (2018). However, with several wind and solar scenarios making the model larger than our deterministic model, if realistic size systems are to be modelled stochastically, then computational efficiency may again be an important avenue for further research.

## Acknowledgments

The authors would like to thank Natural Resource and Engineering Research Council of Canada Discovery Grant for funding this project.

## References

F. Abbaspourtorbati, A.J. Conejo, J. Wang and R. Cherkaoui (2017). Pricing electricity through a stochastic non-convex market-clearing model. *IEEE Transactions on Power Systems*, 32 (2), 1248-1259.

V. Araoz, & K. Jörnsten. (2011). Semi-Lagrangean approach for price discovery in markets with nonconvexities. *European Journal of Operational Research*, 214, 411–417.

K. V. d. Bergh, K. Bruninx, E. Delarue, & W. D'haeseleer. (2016). A unit commitment model formulated as a mixed-integer linear program. *TME working paper, Energy and Envirnoment*.

M. Bjørndal, & K. Jörnsten. (2008). Equilibrium prices supported by dual price functions in markets with non-convexities. *European Journal of Operational Research*, 190, 768–789.

F. S. Bresler. (2014). Energy and ancillary services uplift in PJM. *FERC Uplift Workshop*.

R. Fernández-Blanco, J.M. Arroyo and N. Aguacil (2017). On the solution of revenue- and network-constrained day-ahead market clearing under marginal pricing – Part I: an exact bilevel programming approach. *IEEE Transactions on Power Systems*, 32 (1), 208-219.

J. Fortuny-Amat, & B. McCarl. (1981). A representation and economic interpretation of a two-level programming problem. *The Journal of the Operational Research Society*, 32 (9), 783–792.

J. D. Fuller, & E. Çelebi. (2017). Alternative models for markets with nonconvexities. *European Journal of Operational Research*, 261, 436-449.

S. A. Gabriel, A. J. Conejo, J. D. Fuller, B. F. Hobbs, & C. Ruiz. (2013a). *Complementarity Modeling in Energy Markets.* Springer, (Chapter 11).

S. A. Gabriel, A. J. Conejo, C. Ruiz, & S. A. Siddiqui. (2013b). Solving discretely constrained, mixed linear complementarity problems with applications in energy. *Computers and Operations Research*, 40, 1339–1350.

S. A. Gabriel, S. A. Siddiqui, A. J. Conejo, & C. Ruiz. (2013c) Solving discretely-constrained Nash-Cournot games with an application to power markets. *Networks and Spatial Economics*, 13(3), 307-326.

S. A. Gabriel (2017). Solving discretely constrained mixed complementarity problems using a median function. *Optimization and Engineering.* https://doi.org/10.1007/s11081-016-9341-2

F. D. Galiana, A. L. Motto, & F. Bouffard. (2003). Reconciling social welfare, agent profits, and consumer payments in electricity pools. *IEEE Transactions on Power Systems*, 452–459.

P. R. Gribik, W. W. Hogan, & S. L. Pops. (2007, December). Market-clearing electricity prices and energy uplift. Technical Report, Working paper, *Harvard Electricity Policy Group, Harvard University*.

K. W. Hedman, M. C. Ferris, R. P. O'Neill, E. B. Fisher, & S. S. Oren. (2010). Co-optimization of generation unit commitment and transmission switching with N-1 reliability. *IEEE Transactions on Power Systems*, 1052-1063.

W. W. Hogan, & B. J. Ring. (2003, March). On minimum-uplift pricing for electricity mar- kets. Technical Report, Working paper, *Harvard Electricity Policy Group, Harvard University*.

B. Hua and R. Baldick (2016). A convex primal formulation for convex hull pricing. *IEEE Transactions on Power Systems*, DOI: 10.1109/TPWRS.2016.2637718

Huppmann, D. and Siddiqui, S. (2018). An exact solution method for binary equilibrium problems with compensation and the power market uplift problem. *European Journal of Operational Research*, 266(2), 622-638.

C. D. Jonghe, E. Delarue, W. D'haeseleer, & R. Belmans. (2011, August). Integrating real time pricing into unit commitment programming. *Power Systems Computation Conference edition:17*. Stockholm.

G. Liberopoulos, & P. Andrianesis. (2016). Critical review of pricing schemes in markets with non-convex costs. *Operations Research*, 64 (1), 17-31.

R. P. O'Neill, P. M. Sotkiewicz, B. F. Hobbs, M. H. Rothkopf, & R. Stewart. (2005). Efficient market-clearing prices in markets with non-convexities. *European Journal of Operational Research*, 164, 269–285.

C. Ordoudis, P. Pinson, J. M. Morales González, & M. Zugno. (2016). An updated version of the IEEE RTS 24-bus system for electricity market and power system operation studies. Technical University of Denmark.

H. E. Scarf. (1994). The allocation of resources in the presence of indivisibilities. *Journal of Economic Perspectives*, 8 (4), 111–128.

G. Steeger, S. Rebennack (2015). Strategic bidding for multiple price-maker hydroelectric producers. *IIE Transactions*, 47 (9), 1013-1031.

V. K. Tumuluru, Z. Huang, & D. H. K. Tsang. (2014). Integrating price responsive demand into the unit commitment problem. *IEEE Transactions on Smart Grid.* 5 (6), 2757-2765.

Y. Wang, & L. Li (2015). Time-of-use electricity pricing for industrial customers: A survey of U.S. utilities. *Applied Energy.* 149, 89-103.

L. Wu. (2013). Impact of price-based demand response on market clearing and locational marginal prices. *IET Generation, Transmission & Distribution*, 7 (10), 1087-1095.

Y. Zhan and Q.P. Zhenh. (2018). A multistage decision-dependent stochastic bi-level programming approach for power generation investment expansion planning. *IISE Transaction*.

I. Zoltowska. (2016a). Direct Minimum-Uplift model for pricing pool-based auction with network constraints. *IEEE Transactions on Power Systems.* 31 (4), 2538-2545.

I. Zoltowska (2016b). Demand shifting bids in energy auction with non-convexities and transmission constraints. *Energy Economics*, 53, 17-27.

## Appendix A. Notation

Indices (Sets):

| | |
|---|---|
| $i$ | Generation units, $I$ = 1, 2, ..., I |
| $t$ | Index of time periods, $t$ = 1, 2, ..., T |
| $n$ | Index of nodes, $n$ = 1, 2, ..., N |
| $l$ | Index of transmission lines, $l$ = 1, 2, ... , L |

Sub-sets:

| | |
|---|---|
| $I_{ON}$ ($I_{OFF}$) | Generation units that should be on (off) for a couple of periods following previous horizon due to minimum uptime (downtime) constraint |
| $dn$ ($tn$) | Demand nodes or consumer nodes (transfer nodes or no consumer on these nodes) |

*Parameters:*

| | |
|---|---|
| $\lambda_{it}$ | Marginal cost of generator $i$ at period $t$ |
| $\lambda_{it}^{RES}$ | Marginal cost of reserve capacity of generator $i$ at period $t$ |
| $C_{it}^{SU}$ | Startup cost of generator $i$ at period $t$ |
| $C_{it}^{SD}$ | Shut down cost of generator $i$ at period $t$ |
| $C_{it}^{RP}$ | Ramping cost of generator $i$ at period $t$ |
| $RU_i$ | Maximum ramping up for generator $i$ |
| $RD_i$ | Maximum ramping down for generator $i$ |
| $SU_i$ | Maximum startup rate of generation for generator $i$ |
| $SD_i$ | Maximum shutdown rate for generator $i$ |
| $G_i^{max}$ | Maximum power generation of generator $i$ |
| $G_i^{min}$ | Minimum power generation of generator $i$ |
| $D_i^{min}$ | Minimum down time of generator $i$ |
| $U_i^{min}$ | Minimum up time of generator $i$ |
| $Z_{i0}^{PT}$ | The initial status of generator $i$ equals to operation status of generator at period T of previous horizon |
| $G_{i0}^{PT}$ | The initial status of generator $i$ equals to generation quantity of generator at period $T$ of previous horizon |
| $t_i^{ON}(t_i^{OFF})$ | The amount of periods that generation unit $i$ should be on (off) following the previous horizon |
| $A_i^{RES}$ | A binary value representing the potential spinning reserve generators |
| $\varphi$ | The percentage of demand that is required as spinning reserve at time t |
| $a_{n,t}, b_{n,t}$ | Parameters of price, demand, and benefit functions |
| $\bar{Q}_{n,t}$ | Maximum demand of node $n$ at period $t$ |
| $Q_{n,t}^{min}$ | Minimum demand of node $n$ at period $t$ |
| $P_{n,t}^h$ | Price related to minimum demand of node $n$ at period $t$ |
| $P_{n,t}^l$ | Price related to maximum demand of node $n$ at period $t$ |
| $A_{i,n}^{gen}$ | A binary matrix linking generators to nodes |
| $A_{l,n}^{grid}$ | A matrix with values -1, 0, or 1 linking nodes to transmission lines |
| $F_l^{max}$ | Maximum capacity of power flow on line $l$ |
| $B_l$ | Susceptance of line $l$ |
| $\bar{\theta}$ | voltage angle limit |

*Decision variables:*

| | |
|---|---|
| $z_{it}$ | Binary variable for operation status of generator $i$ at period $t$ |
| $v_{it}$ | Binary variable for startup status of generator $i$ at period $t$ |
| $w_{it}$ | Binary variable for shutdown status of generator $i$ at period $t$ |
| $g_{it}$ | Power generation level of generator $i$ at period $t$ |
| $g_{it}^R$ | Reserve capacity of generator $i$ at period $t$ |
| $q_{n,t}$ | Demand of nodes $n$ at period $t$ |
| $r_{it}$ | Ramping amount of generator $i$ at period $t$ |
| $f_{l,t}$ | Power flow on line $l$ at period $t$ that can be negative or positive depending on direction |
| $\theta_{n,t}$ | State variable of node $n$ at time period $t$ |
| $u_{nt}$ | Binary variable for FAM constraint |

*Dual variables:*

| | |
|---|---|
| $p_{n,t}$ | Dual variable of market clearing constraint (price of node $n$ at period $t$) (4c) |
| $p_t^R$ | Dual variable of reserve constraint (reserve price at period $t$) (4b) |
| $\alpha_{i,t}^{RP1}, \alpha_{i,t}^{RP2}, \alpha_{i,t}^{RP3}, \alpha_{i,t}^{RP4}$ | Dual variables of constraints for ramping (2b-2e) |
| $\alpha_{i,t}^{min}, \alpha_{i,t}^{max}$ | Dual variables of constraints for minimum and maximum level of generation (2f, 2g) |
| $\alpha_{i,t}^{DT}, \alpha_{i,t}^{UT}$ | Dual variables of minimum down and up time constraints (2h, 2i) |
| $\alpha_{i,t}^{OP}$ | Dual variable of operational constraint (2j) |
| $\alpha_{i,t}^{ON}, \alpha_{i,t}^{OFF}$ | Dual variables of must run (off) periods (2k, 2l) |
| $\beta_{n,t}$ | Dual variable of minimum demand constraint (1d) |
| $\gamma_{l,t}^{max}, \gamma_{l,t}^{min}$ | Dual variables of maximum and minimum flow of transmission lines (3b, 3c) |
| $\gamma_{l,t}^{phys}$ | Dual variable of physical constraint using DC power flows (3d) |
| $\gamma_{l,t}^{\theta max}, \gamma_{l,t}^{\theta min}, \gamma_t^1$ | Dual variables of voltage angles constraints (3e, 3f) |
| $\gamma_t^{grid}$ | Dual variable of grid constraint (3g) |
| $\alpha_{i,t}^Z$ | Dual variable of relaxed binary variables (5a) |
| $\sigma_{i,t}$ | Dual variable of extra operational constraint (2q) |